\documentclass[%
	amsmath,
	amssymb,
	aps,
	showpacs,
	showkeys,
	twocolumn,
	altaffilletter,
	nolongbibliography,
	numerical,
	flushbottom,
	secnumarabic,
	pra,
	superscriptaddress,
	floatfix
]{revtex4-1}

\usepackage{lipsum}

\usepackage[dvipsnames]{xcolor}

\usepackage{bm}

\usepackage{graphicx}

\usepackage{upgreek}

\usepackage{natbib}

\usepackage{braket}

\usepackage{physics}

\usepackage{siunitx}

\usepackage[caption=false]{subfig}
\usepackage[%
	breaklinks,%
	pdftex,%
	hyperfootnotes=true,%
	pdfpagelabels,%
	bookmarks,%
	pageanchor,%
]{hyperref}

\hypersetup{%
	colorlinks=true, linktocpage=true, pdfstartpage=1, pdfstartview=FitH, pdfborder={0 0 0},%
	breaklinks=true, pdfpagemode=UseNone, pageanchor=true, pdfpagemode=UseOutlines,%
	plainpages=false, bookmarksnumbered, bookmarksopen=true, bookmarksopenlevel=1,%
	hypertexnames=true, pdfhighlight=/O,
	urlcolor=Red!50!Sepia, linkcolor=NavyBlue, citecolor=Emerald, 
}

\newcommand{\ie}{i.\,e.}
\newcommand{\ped}[1]{_\text{#1}}
\newcommand{\api}[1]{^\text{#1}}

\newcommand{\nspin}{N}
\newcommand{\ham}{H}
\newcommand{\tf}{t\ped{f}}
\newcommand{\diss}{\mathcal{D}}
\newcommand{\partitionFunction}{\mathcal{Z}}

\newcommand{\DELTA}{\mathop{}\!\Updelta}

\DeclareMathOperator{\eu}{e}
\DeclareMathOperator{\iu}{i}

\begin{document}

\author{G.~Passarelli}
\affiliation{Dipartimento di Fisica, Universit\`a di Napoli ``Federico II'', Monte S.~Angelo, I-80126 Napoli, Italy}

\author{G.~De Filippis}
\affiliation{Dipartimento di Fisica, Universit\`a di Napoli ``Federico II'', Monte S.~Angelo, I-80126 Napoli, Italy}
\affiliation{CNR-SPIN, Monte S.~Angelo  via Cinthia,  I-80126 Napoli, Italy}

\author{V.~Cataudella}
\affiliation{Dipartimento di Fisica, Universit\`a di Napoli ``Federico II'', Monte S.~Angelo, I-80126 Napoli, Italy}
\affiliation{CNR-SPIN, Monte S.~Angelo  via Cinthia,  I-80126 Napoli, Italy}

\author{P.~Lucignano}
\affiliation{CNR-SPIN, Monte S.~Angelo  via Cinthia,  I-80126 Napoli, Italy}
\affiliation{Dipartimento di Fisica, Universit\`a di Napoli ``Federico II'', Monte S.~Angelo, I-80126 Napoli, Italy}

\title{A dissipative environment may improve the quantum annealing performances of the ferromagnetic \texorpdfstring{$ \bm{p} $}{p}-spin model}

\date{\today}


\begin{abstract}
We investigate the quantum annealing of the ferromagnetic $ p $-spin model in a dissipative environment ($ p = 5 $ and $ p = 7 $). 
This model, in the large $ p $ limit, codifies the Grover's algorithm for searching in an unsorted database. The dissipative environment is described  by a phonon bath in thermal equilibrium at finite temperature. The dynamics is studied  in the framework of a Lindblad master equation for the reduced density matrix describing only the spins. Exploiting the symmetries of our model Hamiltonian, we can describe many spins and extrapolate expected trends for large $ N $, and $ p $. While at weak system bath coupling the dissipative environment has detrimental effects on the annealing results, we show that in the intermediate coupling regime, the phonon bath seems to speed up the annealing at low temperatures. 
This improvement in the performance is likely not due to thermal fluctuation but rather arises from a correlated spin-bath state and persists even at zero temperature.
This result may pave the way to a new scenario in which, by appropriately engineering the system-bath coupling, one may optimize quantum annealing performances below either the purely quantum or classical limit.
\end{abstract}

\maketitle

\section{Introduction}
\label{sec:motives-qa}

Hard optimization problems can be  mapped onto Ising spin Hamiltonians, whose ground states  (GSs) encode the solution of the given problem~\cite{lucas:np-problems}.  Finding the GS  configuration is then the key issue in many optimization tasks. A well known case is the Ising spin glass~\cite{sherrington-kirkpatrick,parisi:solutions}. A very common strategy to obtain the GS configuration is the so called thermal or simulated annealing (SA)~\cite{kirkpatrick:sa},  where the  main idea  is to ``freeze'' the system in its ground state by slowly reducing its temperature $ T $ towards  zero. Unfortunately, SA, when applied to complex models as the Ising spin glass, can  suffer of a severe slowing down, making the approach unfeasible. 

By contrast, it has been suggested that quantum annealing (QA)~\cite{kadowaki:qa}, employing quantum---rather than thermal---fluctuations, could reduce the slowing down allowing to reach the GS. The QA proceeds from an initial Hamiltonian with a trivial ground state (easy to prepare), to a final Hamiltonian whose ground state encodes the solution of the computational problem. The adiabatic theorem guarantees that the system will track the instantaneous ground state if the Hamiltonian varies sufficiently slowly. That is  why QA is also referred to as adiabatic quantum computation (AQC)~\cite{farhi:quantum-computation}. In the last few years, there has been a renewed interest in QA~\cite{harris:d-wave, ronnow:quantum-speed-up, boixo:experimental-signature,boixo:hundred-qubits,shin:d-wave}. 
 It has been shown that in some cases QA performs better than thermal annealing~\cite{santoro:spin-glass,martonak:salesman}. However, there are also cases where QA performs worse~\cite{battaglia:qa-3-sat, polkovnikov:sa-qa}. To the date, there are only a few problems where this quantum speed-up has been clearly demonstrated~\cite{grover:search}, while in general such a rigorous evidence is missing and one must rely on numerical simulations, with outcomes strongly depending on the specific problem addressed.

Physical implementation of quantum annealers~\cite{Johnson10,Harris10,King15} on a finite number of spins (up to thousand of spins) have been already used to obtain the GS of  complex spin models, but a significant improvement compared to SA has not been yet demonstrated.

In a realistic system, the presence of an unavoidable dissipative environment requires  approaching the problem of QA with great care. Although adiabatic quantum computation has been shown to  be less sensitive to thermal noise with respect to universal quantum computation~\cite{albash:decoherence}, thermal relaxation phenomena, in general, are expected to have a negative effect on quantum adiabatic algorithms, since thermal excitations decrease the probability of finding the system in the lowest-lying energy state and the eigenstates populations  are expected to tend to the Gibbs equilibrium populations after a relaxation time $T_1$~\cite{breuer:open-quantum,levitt:spin}. Exceptions to this behavior have been shown in Ref.'s~\onlinecite{Campagnano:PLA10,Campagnano:NJP08}. Moreover, in a recent paper it has been proven that the working temperature must be appropriately scaled down with the problem size to be confident with the result~\cite{Hen17}. 

In specific cases, however, it has been suggested that the external environment may be even  beneficial in reaching the target ground state showing better performance than closed-system quantum annealing~\cite{amin:thermal-aqc,dickson:thermal-qa,arceci:dissipative-lz,Smelyanskiy:decoherence,Smelyanskiy:decoherence2}. The point here  is that  the evolution of a far-from-thermal equilibrium spin system coupled to a large set of oscillators describing the external environment is not fully understood. The environment is no longer a mere source of decoherence, but can participate to the system dynamics in a non trivial way.

In order to infer the behavior  of  ``realistic''  macroscopic quantum devices, we study a large $ \nspin $ spins system with a reasonably simple, yet non trivial model Hamiltonian, to get sufficiently close to the thermodynamical limit. Since the Hilbert space dimension describing $\nspin$ qubits  grows exponentially (as $ 2^\nspin $), we focus onto a model Hamiltonian having a spin-symmetry that allows us to work with Hilbert spaces of reasonable dimensions,  the so-called ferromagnetic $ p $-spin model~\cite{kirkpatrick:pspin-interaction-spin-glass,jorg:energy-gaps}, which we will introduce in the next section. Then the effect of the environment on the dynamics of such a system is studied comparing what happens with and without the coupling to a set of oscillators that mimics the external environment. The main result of the paper is that, for coupling strong enough, the environment ``helps'' the annealer to reach the target GS in a shorter time. Such speed-up seems to be an open issue, as, in the case of the 1D Ising chain, it has been observed in Ref.~\onlinecite{Smelyanskiy:decoherence}, but not in Ref.~\onlinecite{keck:ising}.  Following Ref.~\onlinecite{Wauters:2017}, we compare our dynamics also with that obtained by the simulated annealing, discussing limitations and advantages of the two approaches.

\section[Ferromagnetic \texorpdfstring{$ p $}{p}-spin model]{Ferromagnetic p-spin model}
\label{sec:motives-p-spin}

The ferromagnetic $ p $-spin model is an Ising spin system in which  each spin interacts  with $ p - 1 $ other spins~\cite{nishimori:non-stoq}. This model is particularly interesting since, in the limit $ p\to\infty $, it codifies the Grover's algorithm for searching in an unsorted database~\cite{grover:search}.  Classical  algorithms require  $ 2^\nspin $ steps  (where $ \nspin $ is the number of entries) to solve such a problem.  However  quantum mechanics allows for a quadratic speed-up (\ie, $ 2^{\nspin/2} $ steps are required)~\cite{grover:search}. As mentioned in the introduction, we focused on this model because it allows to study larger systems exploiting the total spin conservation.

The $ p $-spin Hamiltonian is given by:
\begin{equation}\label{eq:motives-p-spin-quantum}
	\ham\ped{p} = -\nspin \qty(\frac{1}{\nspin} \sum_{i=1}^{\nspin} \sigma^z_i)^p.
\end{equation}
where  the Pauli matrix $ \sigma^z_i $ refers to the $i\api{th}$ spin.
Quantum fluctuations are introduced by  a transverse field:
\begin{equation}\label{eq:motives:transverse-field}
	\ham_0 = -\Gamma\sum_{i=1}^{\nspin} \sigma^x_i, 
\end{equation}
and the full time-dependent Hamiltonian is built as a linear interpolation between~\eqref{eq:motives-p-spin-quantum} and~\eqref{eq:motives:transverse-field}:
\begin{equation}\label{eq:motives-hamiltonian}
	\ham(t) = \qty(1-\frac{t}{\tf}) \ham_0 + \frac{t}{\tf} \ham\ped{p}.
\end{equation}

The linear schedule is the simplest possible one, yet other interpolating functions may be tested~\cite{avron:optimal-time-schedule}. 
We choose $ \Gamma $ as our reference energy scale  (and $\tau=\hslash/\Gamma$ as time scale) except where explicitly mentioned. 
The evolution of the system state $ \ket{\psi} $ is evaluated by means of a dynamical equation for the corresponding density matrix $ \rho = \dyad{\psi} $, in the presence of a dissipative bath made up of harmonic oscillators (phonons)~\cite{caldeira:caldeira-leggett}. This dynamical equation is known as Lindblad master equation~\cite{breuer:open-quantum,zanardi:master-equations}, and it guarantees the complete positivity of the density matrix at any time, hence preserving the probabilistic interpretation of its diagonal elements in the Hamiltonian eigenbasis~\cite{breuer:open-quantum,zanardi:master-equations}. It reads as
\begin{equation}\label{eq:motives-lindblad}
	\dv{\rho(t)}{t} = -\iu{} \comm\big{\ham + \ham\ped{LS}}{\rho(t)} + \diss\qty\big[\vphantom{()}\rho(t)],
\end{equation}
where $ \ham\ped{LS} $ is the Lamb shift (LS) Hamiltonian and $ \diss $ is the dissipator super-operator (see appendix~\ref{app:lindblad}). These terms appear because of the coupling with the environment~\cite{breuer:open-quantum}.
In deriving equation~\eqref{eq:motives-lindblad}, we assume that the thermal bath is in an equilibrium state at an inverse temperature $ \beta $, and that system-bath correlations can be disregarded because of small system-bath couplings ({Born approximation}); moreover, the evolved density operator does not have memory of itself at preceding times ({Markov approximation}) and is calculated within the rotating wave approximation, which enforces the energy conservation. 

\section{Annealing procedure}
\label{sec:motives-annealing-procedures}

At $ t = 0 $, the annealing starts by preparing the system in the trivial ground state of the Hamiltonian~\eqref{eq:motives:transverse-field}. In the $ \sigma^z $ basis, also called computational basis, it reads as
\begin{equation}
	\ket{\psi(t = 0)} = \bigotimes_{i=1}^{\nspin} \qty[\frac{1}{\sqrt{2}} \qty(\ket{0}_i^{\vphantom{j}} + \ket{1}_i)],
\end{equation}
and quantum fluctuations continuously flip each spin from up to down (and vice versa) at a rate $ \Gamma /\hslash $.
The full system Hamiltonian~\eqref{eq:motives-hamiltonian} commutes with the total spin operator $ S^2 $. Both the initial and the final state belong to the subspace with the largest eigenvalue of $ S^2 $, thus the dynamics will never bring the evolved ket state outside this subspace. Also the coupling to the environment, that will be introduced in the following, preserves this  property. Hence, instead of studying the full Hilbert space of dimension $ 2^\nspin $, we can restrict our analysis to the eigenspace associated with $ S = \nspin/2 $, which has dimension $ N + 1 $. This provides an exponential simplification in studying the behavior  of this system in the large $ N $ limit.

The strength of the transverse field is then progressively reduced to zero in a time $ \tf $. The effectiveness of the annealing is quantified by calculating some relevant observables, such as the fidelity, that is, the probability of finding the system in the ground state, and the residual energy, that is the difference between the exact ground state energy of the $ p $-spin Hamiltonian, and the instantaneous energy at $ t = \tf $. The latter is a powerful indicator if one is interested in finding just one configuration that minimizes the Hamiltonian~\eqref{eq:motives-p-spin-quantum} without having to concern about accidental degeneracies:
\begin{equation}\label{eq:motives-residual-energy}
	\epsilon\ped{res}(\tf) = \frac{1}{\nspin} \qty(\ev*{\ham\ped{p}} - E\ped{GS});
\end{equation}
$ E\ped{GS} $ is the target GS energy.
The adiabatic theorem of quantum mechanics ensures that if the evolution is slow enough ($ \tf\to\infty $) the system will remain in its instantaneous ground state at any time~\cite{born:adiabatic-theorem}, hence we expect the residual energy to decrease to zero with increasing $ \tf $. The optimal $ \tf $ has to be larger than  the inverse of the squared minimum gap $ \Delta $ between the ground state and the first excited state~\cite{kato:adiabatic,amin:adiabatic,jensen:adiabatic}. This means that we expect the residual energy to scale as $ \tf^{-2} $ when a fully adiabatic regime is reached. 
Indeed, if the annealing time is too short, a succession of diabatic Landau-Zener (LZ) transitions will excite the system and reduce the fidelity of the adiabatic algorithm~\cite{zener:crossings}.

\section{Quantum annealing without coupling to the environment}
\label{sec:motives-qpt}

In this section we shall describe the annealing of the isolated system at $ T = 0 $ to be compared with that of the open system described in the following.

The adiabatic theorem limits the applicability of AQC to systems with non-vanishing energy gaps. 
The $ p $-spin ferromagnetic model is subject to a quantum phase transition (QPT) at $ T = 0 $, separating a disordered paramagnetic phase to an ordered ferromagnetic phase. At the quantum critical point, the minimum gap $ \Delta $ approaches to zero in the thermodynamical limit, and the annealing time required to satisfy the adiabatic theorem diverges~\cite{sachdev:quantum-phase-transitions}.  

For $ p = 2 $, the $ p $-spin Hamiltonian has a second-order QPT, and the minimum gap scales as~\cite{bapst:quantum-spin-glass} $ \Delta \sim \nspin^{-1/3} $. By contrast, for  $ p > 2 $, the model has a first-order QPT, with an exponentially vanishing minimum gap~\cite{bapst:quantum-spin-glass}  when $ \nspin\to\infty $.

\begin{figure}[tb]
	\centering
	\includegraphics[width=0.95\linewidth]{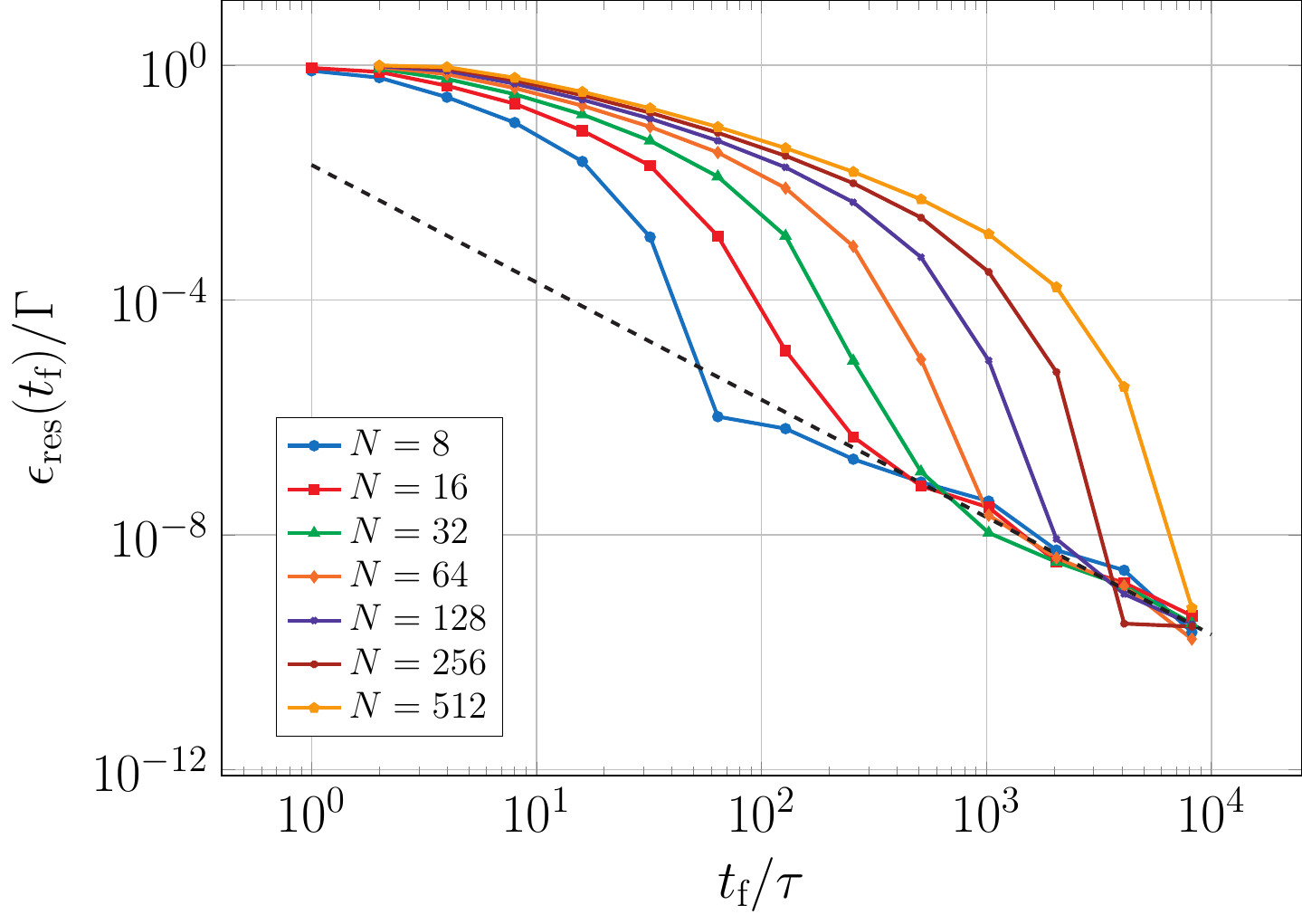}
	\caption{Residual energy in units $ \Gamma $ as a function of the annealing time $ \tf $ in units $ \tau $, for the Hamiltonian~\eqref{eq:motives-hamiltonian} with $ p = 2 $ (bilogarithmic scale). Three different regimes can be observed: a constant beginning region, an intermediate LZ region and the final power-law tail proportional to $ 1/\tf^2 $.}
	\label{fig:motives-p-2-residual-close}
\end{figure}

\begin{figure}[tb]
	\centering
	\includegraphics[width=0.95\linewidth]{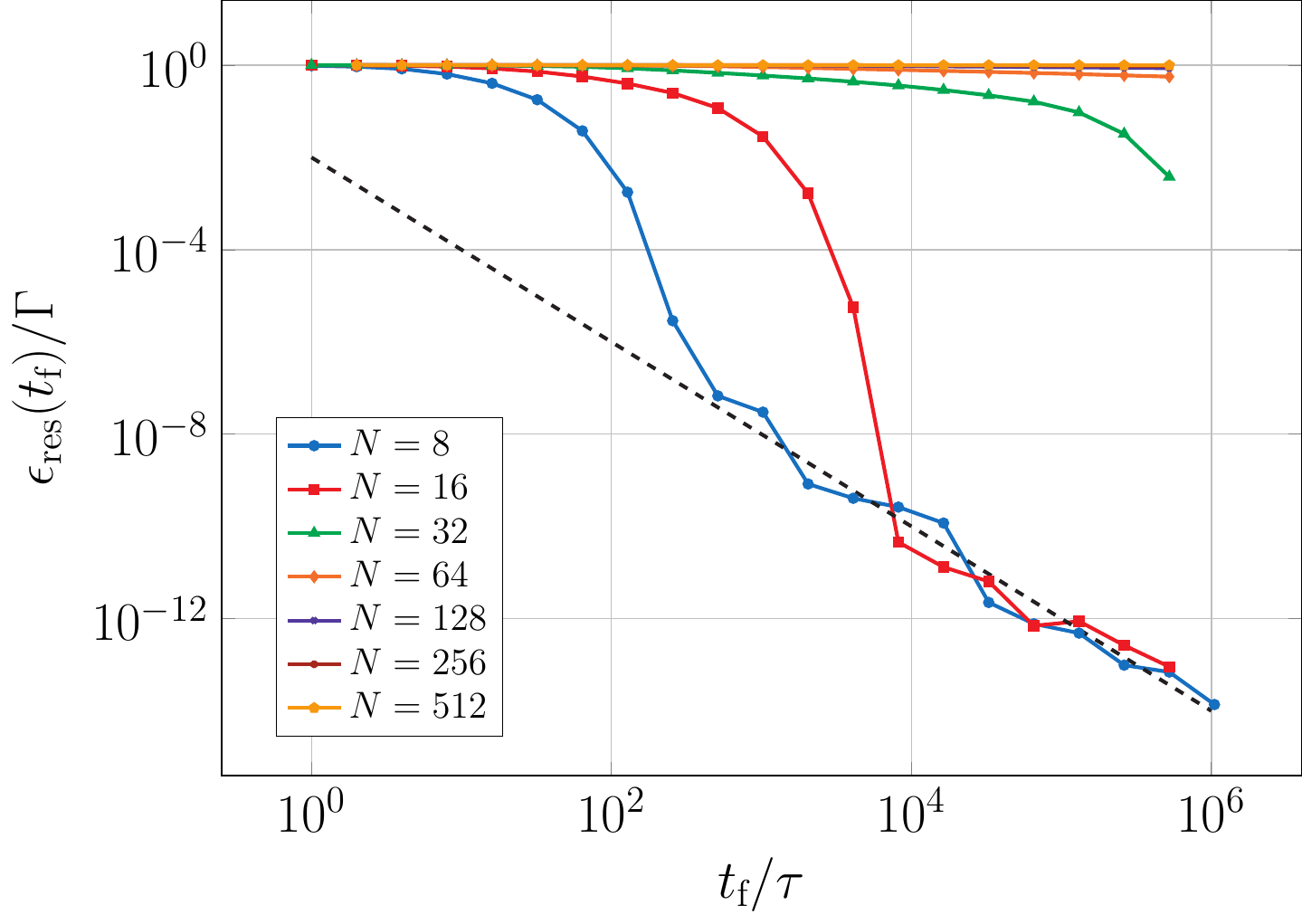}
	\caption{Residual energy in units $ \Gamma $ as a function of the annealing time $ \tf $ in units $ \tau $, for the Hamiltonian~\eqref{eq:motives-hamiltonian} with $ p = 5 $ (bilogarithmic scale). The power-law adiabatic tail is not visible when $ \nspin > 16 $ in the analyzed range of annealing times.}
	\label{fig:motives-p-5-residual-close}
\end{figure}

The behavior of the residual energy as a function of $ \tf $ reflects the gap dependence on the system dimension. As an example, in figure~\ref{fig:motives-p-2-residual-close} we show the behavior of the residual energy as a function of the total annealing time for $ p = 2 $, and for various dimensions of the spin chain $ \nspin $. The curves in figure~\ref{fig:motives-p-2-residual-close} show three different regimes. In the first regime, the system state remains trapped in the paramagnetic phase. The annealing time is too short for the system to follow the ground state across the critical point, and the residual energy is approximately constant~\cite{Wauters:2017}. In the third regime, the residual energy scales as $ 1/\tf^2 $ independently from the system size, as predicted by the adiabatic theorem. The intermediate regime is governed by the diabatic Landau-Zener transitions~\cite{zener:crossings}. This suggests that, in the intermediate regime, the residual energy scales as
\begin{equation}
	\epsilon\ped{res} (\tf) = \frac{C}{\nspin} \eu^{-\tf/\tau_\nspin},
\end{equation}
where $ C $ is a dimensional constant and $ \tau_\nspin $ is proportional to $ \Delta^{-2} $, hence depends on $ \nspin^{2/3} $. Thus, the larger is $ \nspin $, the larger is the time $ \tf $ needed to satisfy the adiabatic theorem~\cite{Wauters:2017}.

When $ p = 5 $, the residual energy behaves as shown in figure~\ref{fig:motives-p-5-residual-close}. The first and the third regimes are very similar to that for the case $ p = 2 $. By contrast, the intermediate regime is different, because in this case the minimum gap exponentially vanishes in $ \nspin $, hence the characteristic time of the LZ transitions increases exponentially~\cite{Wauters:2017}. 

\section{Quantum annealing with decoherence}
\label{sec:motives-finite-temperature}

At finite temperatures $ T \neq 0 $, the $ p $-spin system is subject to a classical phase transition (CPT). The critical temperature $ T\ped{c} $ separates the ordered ferromagnetic phase ($ T < T\ped{c} $) from the disordered paramagnetic phase ($ T > T\ped{c} $).
In addition,  thermal excitations tend to populate excited states. This effect is relevant when the temperature $ T $ is comparable or larger than the minimum gap $ \Delta $. Thus, we expect the fidelity to approach the Boltzmann equilibrium value for long $ \tf \gg T_1 $
\begin{equation}
	P_1\api{eq}(\tf) = \frac{\eu^{-\beta E_1(\tf)}}{\partitionFunction},
\end{equation}
where  $\partitionFunction$ is partition function:
\begin{equation}
	\partitionFunction = \sum_{i = 1}^{\nspin+1} \eu^{-\beta E_i}.
\end{equation}

\begin{figure*}[tb]
	\centering
	\subfloat[]{\label{fig:motives-p-5-n-8-open-close-beta-2}\includegraphics[width=0.475\linewidth]{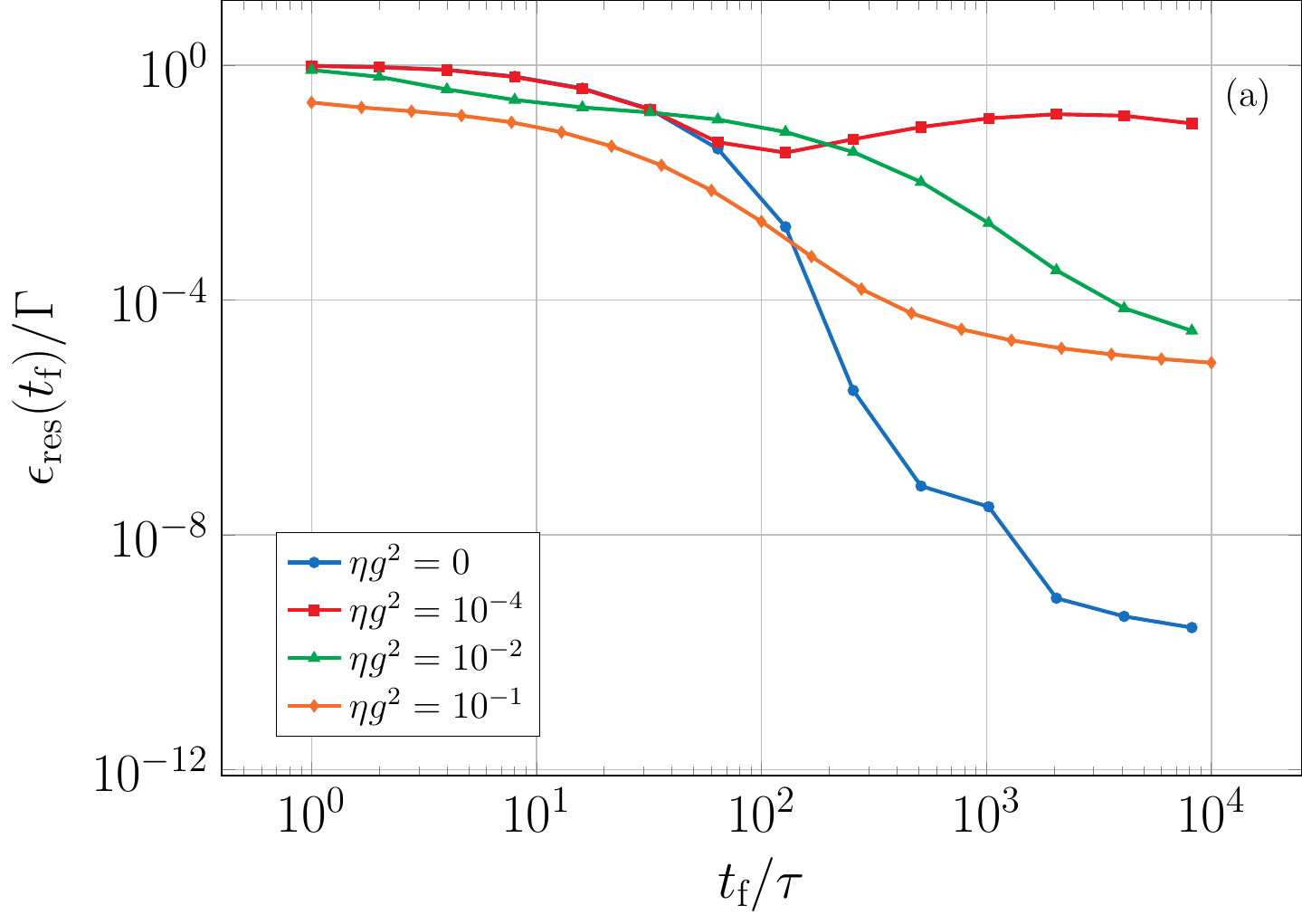}}\hfill
	\subfloat[]{\label{fig:motives-p-5-n-8-open-close-beta-10}\includegraphics[width=0.475\linewidth]{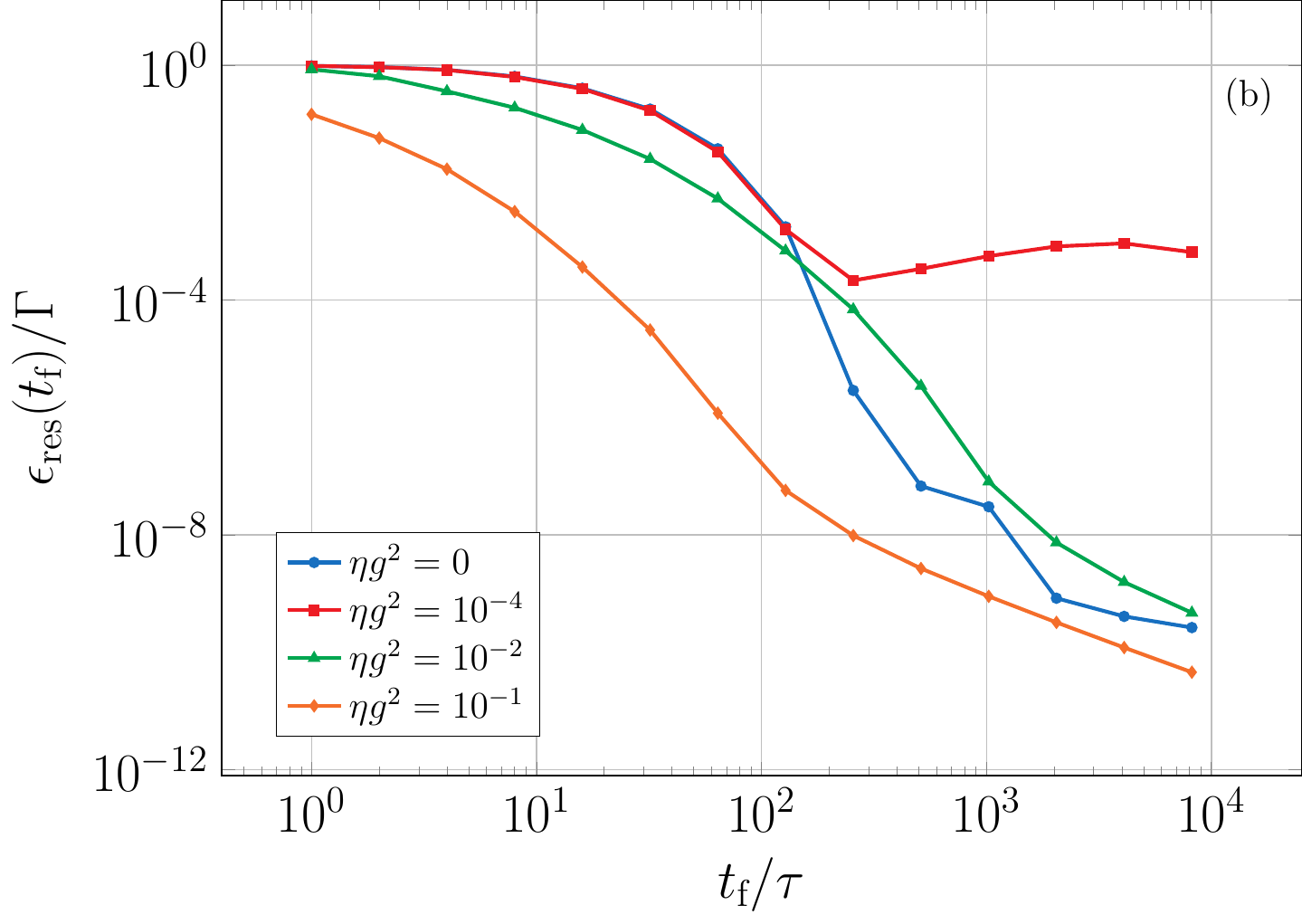}}
	\caption[$ \epsilon\ped{res}(\tf) $, QA (closed and open system, $ p = 5 $)]{Residual energy in units $ \Gamma $ for an open or closed quantum system as a function of the annealing time $ \tf $ in units $ \tau $, for $ \nspin = 8 $ and $ p = 5 $, with $ \beta = 2 $ (panel (a)) and $ \beta = 10 $ (panel (b)). The scale is bilogarithmic. Different system-bath couplings are shown. When the temperature is lower, the driving force of the bath towards the ground state is more evident.}
	\label{fig:motives-p-5-n-8-open-close}
\end{figure*}

In the Lindblad approach, each mode of the thermal bath is coupled to the qubit system through a spectral density function, which is proportional to a coupling energy $ \eta g^2 $ and represents how each phononic mode is coupled to the reduced system (the explicit form of the coupling is described in appendix~\ref{app:lindblad}). 

%
%
In  this section we shall focus on small chains ($ 8 $ sites) because of the large computational cost  of building the dissipator $ \diss $ and the Lamb shift Hamiltonian at each time step.
We choose $p=5$ to study the hard case in which the quantum phase transition of our model is first-order.  

In figure~\ref{fig:motives-p-5-n-8-open-close}, we compare the residual energy of a closed system ($\eta g^2=0$) with $ \nspin = 8 $ and $ p = 5 $ to that of an open system, coupled with $ \eta g^2 = 10^{-4} $ or $ \eta g^2 = 10^{-2} $ to a thermal bath in equilibrium at an inverse temperature $ \beta = 2 $ (left panel) or $ \beta = 10 $ (right panel). 
We choose these two temperatures as they  characterize two different experimentally accessible regimes.
In order to understand what are the  temperatures related to these values we have to restore the real units.
Here all the energies are measured in units of $\Gamma$. We are interested in describing experimental facts related to the current  technology based on superconducting flux qubits \cite{}, where $\Gamma/\hslash$ is of the order of $ \si{\giga\hertz} $, hence  we  fix $\Gamma/\hslash=\SI{1}{\giga\hertz}$. Thus  $\beta=2$ corresponds to $T\sim \SI{25}{\kelvin} $ and $\beta=10$ corresponds to $T\sim \SI{5}{\kelvin}$, describing two interesting (low and ultra-low) operating temperatures, both accessible with currently available dilution fridges.

Inspection of figure~\ref{fig:motives-p-5-n-8-open-close} shows that the effect of the bath is negligible for small $ \tf $ and becomes relevant at longer $ \tf $. 
At intermediate temperatures $\beta=2$, independently of the coupling strength, the unitary dynamics ($\eta g^2=0$) is always more efficient in reaching the ground state.

At low temperatures $\beta=10$, this picture is no longer valid and the scenario becomes richer and more interesting.
By inspection of figure~\ref{fig:motives-p-5-n-8-open-close-beta-10} we can notice that at weak coupling $\eta g^2=10^{-4}$  the bath has a detrimental effect on the annealing procedure and the residual energy (red curve with square dots) is always larger than that of the closed system  (blue curve with circle dots), independently of $\tf$. This is a manifestation of thermalization processes. Unexpectedly enough, by increasing the system-bath coupling, things change drastically.  The residual energy at $\eta g^2=10^{-2}$ (green curve with triangle dots)  is smaller than that of the closed system, until $\tf \sim 10^2$. Further increasing the coupling, at  $\eta g^2=10^{-1}$, the residual energy is always way smaller than that of the isolated system.

These results show that the velocity of convergence to the ground state at low temperatures (with respect to energy gaps) is strongly influenced by the system bath-coupling whose increase  seems to speed up the calculation giving rise to a residual energy that decreases more  and more rapidly to zero.
In particular, the stronger is the coupling, the faster the residual energy goes to zero. 
It is important to say that our results at $\eta g^2=10^{-1}$ may not be as accurate as for weaker couplings.  Indeed  $\eta g^2=10^{-1}$  falls very close to the maximum system bath-coupling  where the Lindblad approach (that is a weak coupling theory) can be applied. However,  we are currently approaching the same problem using a variational approach \cite{workinprogress} and preliminary results seems to confirm this scenario.

\begin{figure}[tb]
	\centering
	\includegraphics[width=0.95\linewidth]{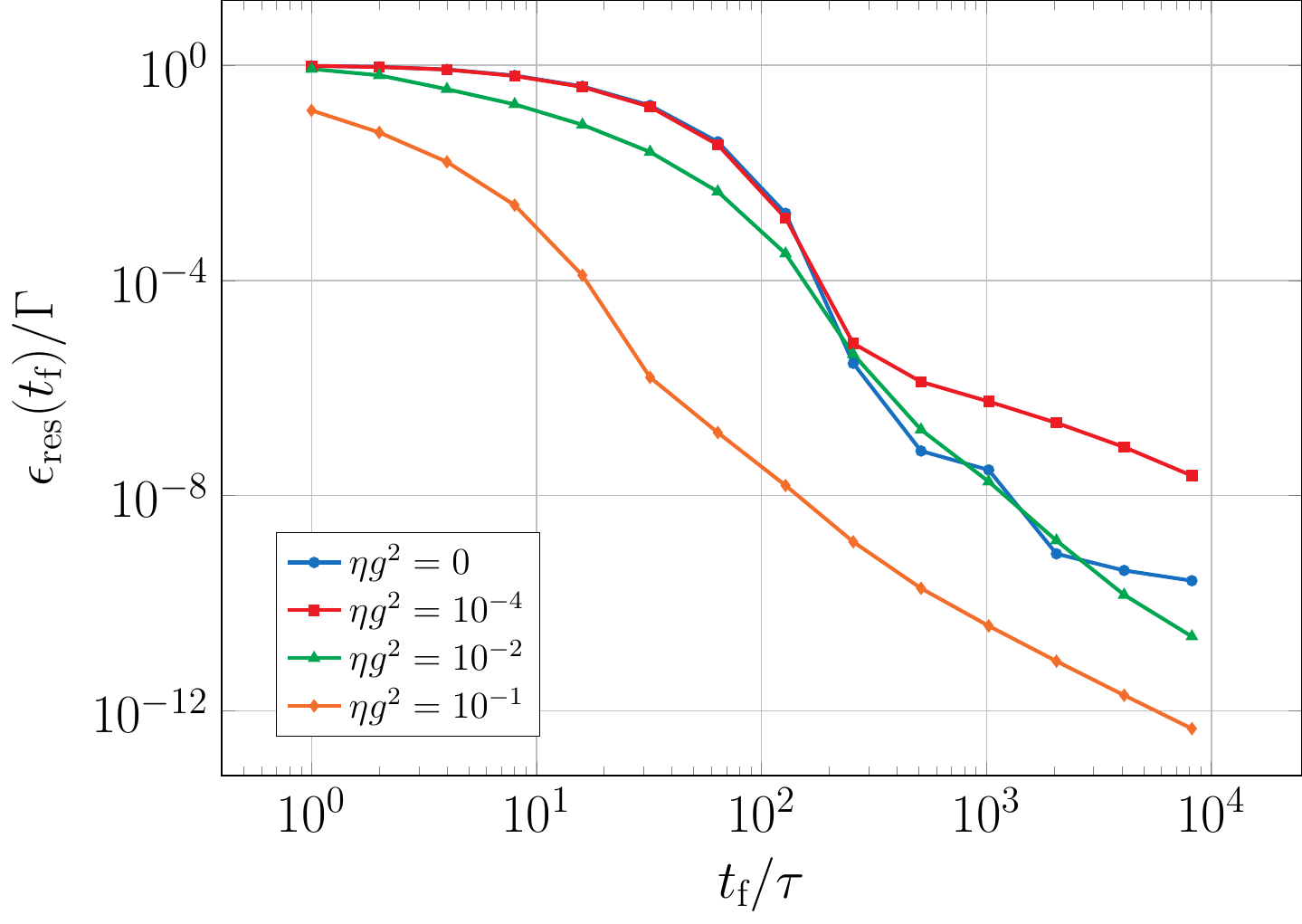}
	\caption{Bilogarithmic plot of the residual energy in units $ \Gamma $ as a function of $\tf$ in units $ \tau $, for $\nspin = 8$, $p = 5$ and $\beta \to  \infty$. The speed-up in the quantum annealing of the open $ p $-spin model is observed also at $ T = 0 $: it is most likely a quantum effect.}
	\label{fig:p-5-n-8-open-closeT-0}
\end{figure}

This  speed-up could either be due to quantum  or  classical effects, or to a combination of both. However, we  guess that it is most likely a quantum effect, as it happens also at $T = 0 $ as clear from figure~\ref{fig:p-5-n-8-open-closeT-0}, and arises because of the formation of an entangled system-bath state that will be addressed in detail in a future publication \cite{workinprogress}.

\section{Quantum vs Thermal annealing}
In this section, we will compare the quantum annealing (both unitary and dissipative) of the $ p $-spin model with the simulated thermal annealing~\cite{kirkpatrick:sa}.
Simulated annealing is performed by linearly reducing the temperature $T(t)=T_0 (1-t/\tf)+T\ped{f}$ from an initial temperature  $T_0$ larger than the critical temperature of the system $T\ped{c}$ to a final temperature $T\ped{f}\ll T\ped{c}$.  
Following Ref.~\onlinecite{Wauters:2017}, simulated annealing is performed using a Glauber master equation for the magnetization of the system, choosing a heat bath form for the transition rates. 

To make a fair comparison between  SA and QA we fix the final temperature of the simulated annealing $ T\ped{f} = 1 / \beta $, where $ \beta $ is the inverse temperature of the phononic bath of the quantum annealing. 
The outcomes of simulated annealing are largely independent  $T_0$ hence we choose $T_0=2$ in all the calculations.

As  shown in Ref.~\onlinecite{Wauters:2017}, for $p=2$  simulated annealing outperforms quantum annealing. This result is mostly due the fact that  the simulated annealing residual energy  decrease exponentially in time and  is independent of the system size. 

The comparison for $ p > 2 $ is less simple, as the residual energy in simulated annealing is no longer  size-independent, and moreover it is more difficult to extrapolate its limiting behavior for large $ \nspin $~\cite{Wauters:2017}.
Due to our difficulties in simulating large systems,  in what follows we focus on $ N = 8 $ and $ p = 5, 7 $.

\begin{figure*}[tb]
	\centering
	\subfloat[]{\label{fig:motives-p-5-sa-qa}\includegraphics[width=0.475\linewidth]{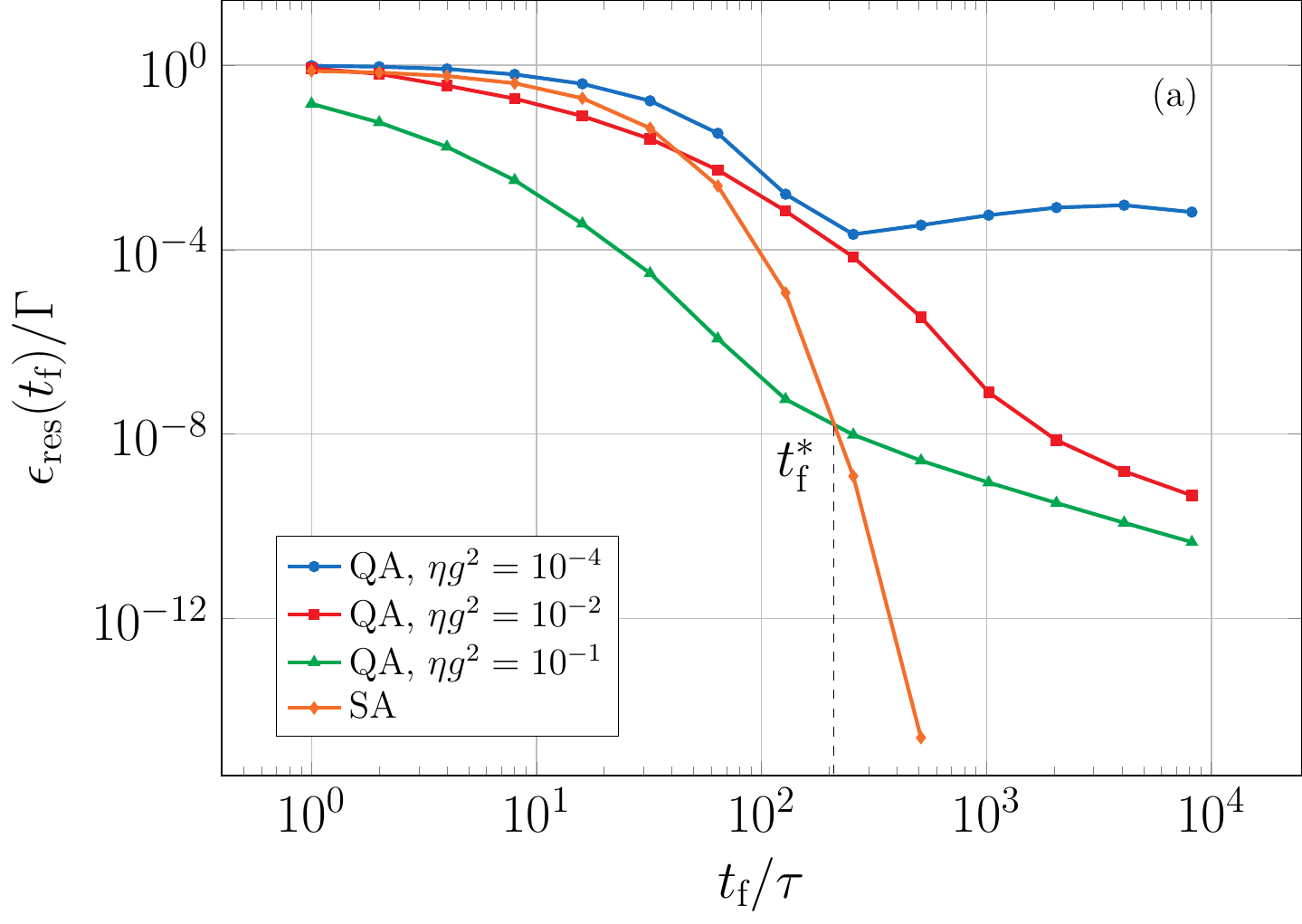}}\hfill
	\subfloat[]{\label{fig:motives-p-7-sa-qa}\includegraphics[width=0.475\linewidth]{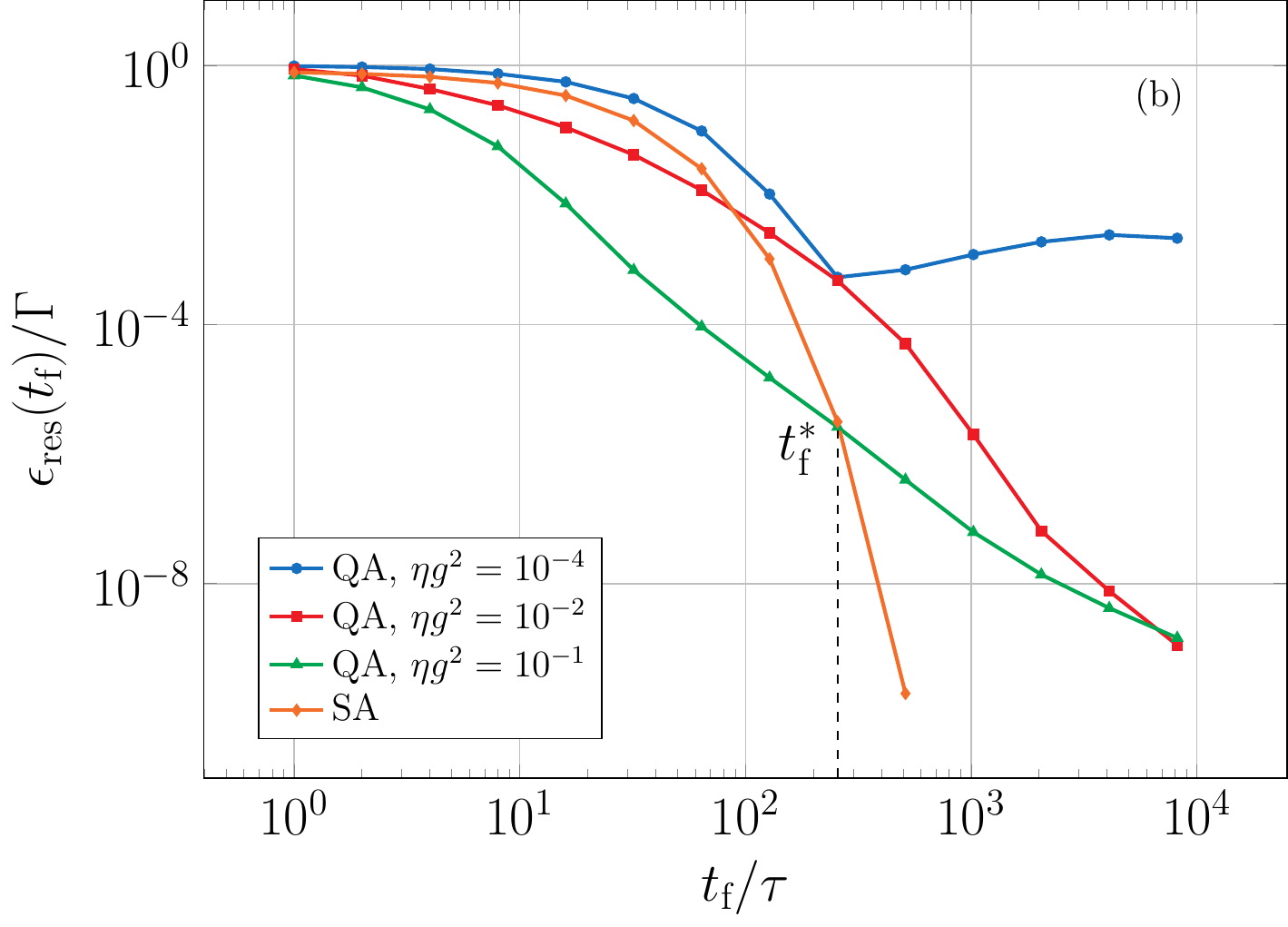}}
	\caption{Comparison of the SA residual energy of a chain of $ \nspin = 8 $ qubits with that at the end of a quantum annealing, for $ \beta = 1/T\ped{f} = 10 $ and $ p = 5 $ (panel (a)) or $ p = 7 $ (panel (b)). The scale is bilogarithmic. Several system-bath couplings are shown for the QA dynamics; a cross-over $ \tf^* $ is present, when thermal annealing starts to outperform QA. Interestingly, $ \tf^* $ seems to become longer with increasing $ p $.}
	\label{fig:motives-p-5-p-7}
\end{figure*}

In the previous section we have shown that at very low temperature ($\beta=10$) and for strong system-bath couplings $\eta g^2 = 10^{-1}$  the environment may help reducing QA residual energy. However, at such temperatures SA is still expected to perform better than QA. The adiabatic theorem ensures a $ 1/\tf^2 $ asymptotic dependence of the residual energy in QA, as opposed to the expected $ 1/\tf $ asymptotic behavior for SA, and this should endorse quantum over thermal annealing for long $ \tf$. However, the minimal time at which the adiabatic regime is recovered is an exponential function of $ \nspin $. Thus, for macroscopic systems the asymptotic behaviors might be reached only at impractically long annealing times, hence the performances of the two techniques have to be compared in the intermediate-$ \tf $ regime. 
At intermediate $ \tf $, the QA of the open system seems to perform better than SA at low temperatures and for strong system-bath coupling (see for example figure~\ref{fig:motives-p-5-sa-qa} for $ \nspin = 8 $, $ p = 5 $ and $ \beta = 10 $). The time $ \tf $ at which simulated annealing starts to outperform quantum annealing seems to be directly proportional to the exponent $ p $, as is evident by comparing figure~\ref{fig:motives-p-5-sa-qa} and figure \ref{fig:motives-p-7-sa-qa}, where we reported our simulations relative to the case $ p = 7 $. However this conclusion necessitates a deeper analysis for longer chains.

\section{Conclusions}
\label{sec:motives-conclusions}

Adiabatic quantum computation (also quantum annealing) is a modern tool employing quantum mechanics to solve a class of optimization problems even if  there is no general consensus on whether or not it can perform faster than conventional computing. From a theoretical of view, quantum annealing has a serious limitation when dealing with systems showing a quantum phase transition, since the effectiveness of an adiabatic algorithm is proportional to the inverse minimum gap in the energy spectrum. In these cases, simulated thermal annealing might be better suited to investigate such systems, as suggested by our simulations.

The $ p $-spin ferromagnetic model, discussed in this paper, shows a second-order QPT for $ p = 2 $ and  a first-order QPT for $ p > 2 $. 

When $ p = 2 $, the minimum energy gap scales as $ \nspin^{-1/3} $ and quantum annealing converges to the ground state in polynomial time;
when $ T\ped{f} = 0 $, the thermal annealing residual energy vanishes exponentially with $ \tf $ and is size-independent, making simulated annealing the method of choice. 

This conclusion cannot be extended trivially to the case $ p > 2 $ where the quantum annealer scaling is much more difficult to obtain. In this case we are not able to provide a definite answer in choosing the faster method among the two of them.

At low temperatures, the thermal bath may speed-up quantum annealing, but simulated thermal annealing is still expected to be faster for long annealing times $ \tf $, because in quantum annealing the residual energy scales as a power-law of the annealing time in the adiabatic regime, opposed to the exponential decrease of the residual energy in SA. However, for intermediate $ \tf $, the out of equilibrium dynamics of the set of oscillators simulating the external environment pushes the interacting $ p $-spin system towards the target GS, providing a faster convergence.  This is achieved until a cross-over time $ \tf^* $ is reached. Unexpectedly, our analysis shows that that $ \tf $  grows with increasing $ p $. This could suggest that for very large $ p $, QA could perform better than SA, in an accessible time window in presence of a realistic (\ie, not extremely weak) coupling to the environment. This effect is likely not due to thermal fluctuations, but rather arises because of a renormalization of the quantum $ p $-spin Hamiltonian for the effect of the bath. For $ p\to\infty $, this may indicate that quantum annealing is faster than thermal annealing when studying the Grover's problem; further analysis for longer chains is needed to test our hypothesis.
\begin{acknowledgments}
We acknowledge enlightening discussions with G.E. Santoro and A. Tagliacozzo.
\end{acknowledgments}

\appendix

\begin{widetext}

\section{Lindblad equation}
\label{app:lindblad}
The equation of motion for the reduced density matrix (representing only the spin variables) used in this work is equation~\eqref{eq:motives-lindblad}, reported here for convenience:
\begin{equation}
\label{eq:lindblad-omega}
	\dv{\rho(t)}{t} = -\iu \comm\big{\ham \vphantom{()} (t) + \ham\ped{LS}(t)}{ \rho (t)} + \diss\qty\big[ \vphantom{()} \rho(t)],
\end{equation}
where the adiabatic dissipator is
\begin{equation}
\label{eq:lindblad-superoperators-omega}
	\diss\qty\big[ \vphantom{()} \rho(t)] = \sum_{\alpha\beta} \sum_{\omega} \gamma_{\alpha\beta} (\omega) \qty[L_{\beta \omega}  (t) \rho  (t) L_{\alpha\omega}^\dagger (t) - \frac{1}{2} \acomm{L_{\alpha \omega}^\dagger (t) L_{\beta\omega}  (t)}{ \rho (t)}]
\end{equation}
and the Lamb shift Hamiltonian takes the form
\begin{equation}
\label{eq:lamb-shift-omega}
	\ham\ped{LS}  (t) = \sum_{\alpha\beta} \sum_{\omega} S_{\alpha\beta}  (\omega) L_{\alpha\omega}^\dagger (t) L_{\beta\omega} (t).
\end{equation}
They are both expressed in terms of the Lindblad operators $L_{\alpha\omega}(t)$, which are defined in the instantaneous energy eigenbasis $ \Set{\epsilon_a(t)} $ as
\begin{equation}\label{eq:lindblad-operators}
	L_{\alpha\omega}(t) = \sum_{\epsilon_a(t) - \epsilon_b(t) = \omega} \ket{\epsilon_a(t)} \mel{\epsilon_a(t)}{A_\alpha}{\epsilon_\beta(t)}\bra{\epsilon_b(t)}.
\end{equation}
The operators $ A_\alpha $ are the spin operators appearing in the general form of the system-bath coupling Hamiltonian
\begin{equation}
	\ham\ped{I} = \sum_{\alpha} A_\alpha \otimes B_\alpha,
\end{equation}
where $ B_\alpha $ are bath operators.

The matrices $\gamma_{\alpha\beta} (\omega)$ and $S_{\alpha\beta} (\omega)$ are respectively the real and imaginary part of the $\Gamma_{\alpha\beta}(\omega)$:
\begin{equation}
\label{eq:spectral-density-decomposition}
	\Gamma_{\alpha\beta}(\omega) = \frac{1}{2} \gamma_{\alpha\beta} (\omega) + \iu S_{\alpha\beta} (\omega),
\end{equation}
that is the Fourier transform of the two point correlation function of the bath 	$\mathcal{B}_{\alpha\beta} (\tau) \equiv \expval{B_\alpha (\tau) B_\beta (0)}$:
\begin{equation}
\label{eq:spectral-density}
	\Gamma_{\alpha\beta} (\omega) \equiv \int_{0}^{\infty} \eu^{\iu \omega \tau} \mathcal{B}_{\alpha\beta} (\tau) \dd{\tau}.
\end{equation}
We suppose that the spin system is coupled to a bath of harmonic oscillators (phonons), described by the Hamiltonian
\begin{equation}
\label{eq:phononic-bath}
	\ham_B = \sum_{k=0}^\infty \omega_k b_k^\dagger b_k,
\end{equation}
where $ b_k $  satisfy the following algebra: $\comm*{b_k^{\phantom{\dagger}}}{b_{k'}} = 0$, $\comm*{b_k^\dagger}{b_{k'}^\dagger} = 0$, $ \comm*{b_k}{b_{k'}^\dagger} = \delta_{kk'}$.
In this work, the interaction Hamiltonian has the form
\begin{equation}
	\ham\ped{I} = \sum_{i=1}^{\nspin} \sigma_i^z \otimes B,
\end{equation}
where the operator $ B $ is expressed in terms of annihilation and creation operators of each phonon mode:
\begin{equation}
	B = g \sum_k \qty(b_k^\dagger + b_k).
\end{equation}
The constant $ g $ couples the $ z $-component of the total spin operator with each mode of the environment, as customary in the spin-boson model~\cite{zanardi:master-equations,breuer:open-quantum,caldeira:caldeira-leggett}. Moreover, we assume that the bath frequency spectrum is continuous, and that the bath is in equilibrium at an inverse temperature $ \beta $; thus, its density operator is just
\begin{equation}
	\rho_B = \frac{\eu^{-\beta \ham_B}}{\partitionFunction}.
\end{equation}
The Fourier transform of the bath correlation function can be expressed as:
\begin{equation}
\label{eq:spectral-function-ohmic}
	\gamma(\omega) 
	= \frac{2 \uppi J\bigl(\abs{\omega}\bigr)} {1 - \eu^{-\beta \abs{\omega}}} 
	g^2 \qty(\Theta(\omega) + \eu^{-\beta \abs{\omega}} \Theta(-\omega)),
\end{equation}
where $ \Theta(\pm \omega) $ are Heaviside functions~\cite{zanardi:master-equations}.
The model is fully specified once we assign the explicit form of the function $ J(\omega) $. In this paper, we employ an Ohmic bath~\cite{breuer:open-quantum}, characterized by
\begin{equation}
	J(\omega) = \eta \frac{\omega^\nu}{\omega\ped{c}^{\nu-1}} \eu^{-\omega/\omega\ped{c}}, \quad \text{with $ \nu = 1 $},
\end{equation}
where $ \omega\ped{c} $ is a high-frequency cut-off and $ \eta $ is a dimensional parameter.

\section{Simulated thermal annealing}
\label{app:glauber}
	
In our model, the Glauber master equation  can be written in terms of the probability  $ \mathbb{P}(m, t) $ of observing a magnetization $ m $:
\begin{align}\label{eq:motives-glauber-magnetization}
	\pdv{\mathbb{P}(m, t)}{t} = \frac{\nspin}{2} \sum_{\alpha=\pm} \qty(1 + \alpha m + \frac{2}{\nspin}) W_{m, m + 2\alpha/\nspin} \mathbb{P}\qty(m - \alpha\frac{2}{\nspin})
	-\frac{\nspin}{2} \sum_{\alpha=\pm} (1 + \alpha m) W_{m - 2 \alpha /\nspin, m} \mathbb{P}(m, t).
\end{align}
The element $ W_{m, m \pm 2/\nspin} $ is  the rate for a single spin-flip that we choose in the heat bath form:
\begin{equation}
	W_{a, b} = \frac{\eu^{-\beta \DELTA E_{ab}/2 }}{\eu^{-\beta \DELTA E_{ab}/2} + \eu^{\beta \DELTA E_{ab}/2}}.
\end{equation}
There are four terms in the right-hand side of equation~\eqref{eq:motives-glauber-magnetization}: the first two increase the probability $ \mathbb{P}(m, t) $ because of transitions from the states with a magnetization that differs of $ \pm 2/\nspin $ from $ m $; the last two terms represent the inverse processes.
	
At $ t = 0 $ the system is originally prepared in the equilibrium configuration at some temperature $ T_0 \gg T\ped{c} $ ($ T_0 = 2 $).  Then, the temperature is decreased with a linear schedule in a time $ \tf $ towards a final temperature $ T\ped{f} $, ideally zero.
At the end of the annealing the residual energy is evaluated similarly to QA
\begin{equation}
	\epsilon\ped{res}(\tf) = \frac{1}{\nspin} \qty(\sum_{m} \ham\ped{c}(m) \mathbb{P}(m, \tf) - E\ped{GS}),
\end{equation}
where $ E\ped{GS} $ is the true GS energy.

\end{widetext}


%

\end{document}